\documentclass[aps,prb,twocolumn,floatfix]{revtex4-1}
\usepackage[latin1]{inputenc}
\usepackage[english]{babel}
\usepackage{graphicx}
\usepackage{multirow}
\usepackage{float}
\usepackage{units}
\usepackage{subfigure}
\usepackage[cm]{fullpage}
\usepackage{dcolumn}
\usepackage{amsmath}
\begin{document}
\title{Diffraction and near-zero transmission of flexural phonons at graphene grain boundaries}
\author{Edit E. Helgee}
\affiliation{Department of Applied Physics, Chalmers University of Technology, SE-412 96, G\"oteborg, Sweden}
\author{Andreas Isacsson}
\affiliation{Department of Applied Physics, Chalmers University of Technology, SE-412 96, G\"oteborg, Sweden}

\begin{abstract}
Graphene grain boundaries are known to affect phonon transport and thermal conductivity, suggesting that they may be used to engineer the phononic properties of graphene. Here, the effect of two buckled grain boundaries on long-wavelength flexural acoustic phonons has been investigated as a function of angle of incidence using molecular dynamics. The flexural acoustic mode has been chosen due to its importance to thermal transport. It is found that the transmission through the boundaries is strongly suppressed for incidence angles close to $35^\circ$. Also, the grain boundaries are found to act as diffraction gratings for the phonons.
\end{abstract}

\maketitle

\section{Introduction}
Grain boundaries in graphene have been found to affect the mechanical, electronic and thermal properties of the material\cite{Yazyev2014review,Yazyev2012energy,Zhang2013nonsym,Lee2013strength, Cao2012mechanical,Fei2013phenomena,Yu2011charact,Tsen2012tailoring}. The grain boundaries commonly consist of dislocations, in the form of pentagon-heptagon defect pairs, and cause out-of-plane buckling of the graphene sheet\cite{Coraux2008buckle,Carlsson2011gbstruct,Liu2011gbstruct,Gomez2010RGO,Warner2013ripple,Lehtinen2013buckling,Huang2011TEM,Kim2011gbs}. Recent experimental studies show that dislocations can be introduced into pristine graphene using a focused electron beam\cite{Warner2013ripple,Warner2012disloc,Lehtinen2013buckling,Robertson2012spatial,Hashimoto2004}, suggesting the possibility of adjusting the properties of the material. 

The possibility of manipulating the properties of graphene could be particularly important in applications related to phononics and heat management\cite{Balandin2012phononics, Pop2012review}, where control of the vibrational properties and thermal conductivity of graphene is essential. The effect of grain boundaries on the thermal conductivity of graphene has previously been studied using both molecular dynamics and Greens function methods \cite{Cao2012Kapitza,Cao2012asymmetric,Bagri2011,Lu2012Greens,Serov2013greens,Tan2013ribbon,Liu2014along}. However, out of these studies only Liu et al.\cite{Liu2014along}, who consider transport along the boundary, mention the influence of out-of-plane buckling. Also, these studies give no detailed insight into the scattering processes of specific phonon modes.

In the present study, we investigate the scattering of long-wavelength flexural acoustic phonons at grain boundaries in graphene for several incidence angles using molecular dynamics (MD). This particluar phonon mode was chosen since it is believed to contribute significantly to the thermal conductivity\cite{Seol2010scienceZA,Lindsay2010thermal}. Two grain boundaries are considered in this paper, one with a misorientation angle of $9.4^\circ$ and one with a misorientation angle of $17.9^\circ$. Both grain boundaries display substantial out-of-plane buckling, with a periodic variation in height along the grain boundary due to the distribution of defects. The boundaries are found to act as diffraction gratings for the phonons, and strongly suppressed transmission is also observed for specific angles. In particular, the transmission is as low as 4 \% for incidence angles near $35^\circ$ at both boundaries.

A previous investigation limited to phonons normally incident on the grain boundary showed that the scattering was due almost entirely to the out-of-plane buckling of the boundary\cite{Helgee2014flexural}. Based on this result a continuum mechanical model was constructed, where the grain boundary was modeled as a static out-of-plane displacement. The model showed good agreement with the MD results. Here, we extend this continuum mechanical model to the case of non-normal angle of incidence in order to gain a qualitative understanding of the scattering mechanism. 

\section{Method}
\begin{figure}
  \begin{center}
    \includegraphics[width=\linewidth]{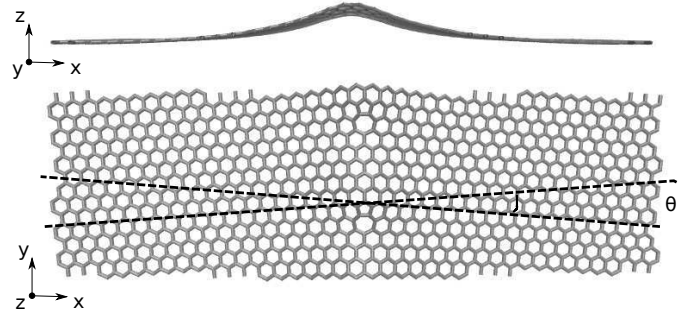}
  \end{center}
  \caption{(Color online) Symmetric tilt grain boundary with misorientation angle $9.4^\circ$, seen from the $y$ direction (top) and $z$ directon (bottom). Figure made using VMD\cite{HUMP96}. \label{fig:GB}}
\end{figure}

All MD simulations have been performed using the program package LAMMPS (large-scale atomic/molecular massively parallel simulator)\cite{Plimpton1995LAMMPS}. The interaction between carbon atoms has been modeled using the Tersoff potential\cite{Tersoff1988Si,Tersoff1988C} with the potential parameters given by Lindsay and Broido\cite{Lindsay2010potential}. This set of parameters has been chosen due to its improved description of acoustic phonon modes in graphene. The considered grain boundaries are symmetric tilt grain boundaries and consists of periodic arrays of pentagon-heptagon defects. The $9.4^\circ$ grain boundary has a period of 1.5 nm in the $y$ direction, parallel to the grain boundary (see Figure \ref{fig:GB}), while the $17.9^\circ$ boundary has a period of 2.4 nm. The grain boundaries have been constructed using the method described in Ref. \citenum{Helgee2014flexural}. For the $9.4^\circ$ boundary this results in a grain boundary buckling $0.6$ nm high and $1.7$ nm wide. Due to the defect distribution the buckling height varies periodically along the grain boundary with an amplitude of $0.06$ nm. The $17.9^\circ$ boundary has a buckling height of 1.5 nm and a buckling width of 5 nm, wih a variation of 0.1 nm along the boundary.

\begin{figure*}
  \begin{center}
    \includegraphics[width=\linewidth]{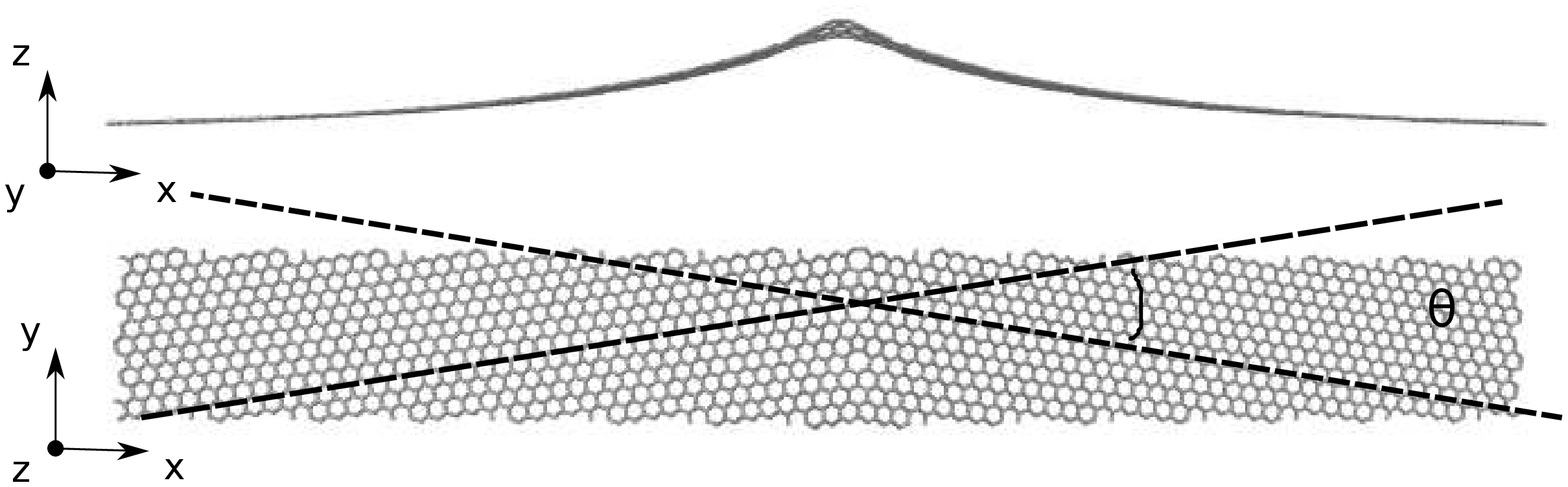}
  \end{center}
  \caption{(Color online) Symmetric tilt grain boundary with misorientation angle $17.9^\circ$, seen from the $y$ direction (top) and $z$ directon (bottom). Figure made using VMD\cite{HUMP96}. \label{fig:GB17_9}}
\end{figure*}

To construct the phonon wavepackets we use the method of Kimmer et al \cite{Kimmer2007WP3}. The displacement $\mathbf{u}_j$ of atom $j$ is then determined by
\begin{equation}
\mathbf{u}_j=\mathrm{Re}\left[\sum_{\mathbf{k}}a_{\mathbf{k}}\vec{\epsilon}_{j\mathbf{k}}e^{i\mathbf{k}\cdot\mathbf{R_j}-i\omega(\mathbf{k})t}\right],
\end{equation}
where $\mathbf{k} = k_x\hat{x}+k_y\hat{y}$ is a wavevector, $\vec{\epsilon}_{j\mathbf{k}}$ is a polarization vector for the considered phonon branch, $\mathbf{R}_j$ is the position vector of atom $j$ and $\omega$ is the phonon frequency. The amplitudes $a_{\mathbf{k}}$ are calculated according to
\begin{equation}
a_{\mathbf{k}}=Ae^{-\eta^2(k_x-k_{0x})^2}e^{-i\mathbf{k}\cdot\mathbf{R}_0},
\end{equation}
where $A$ is an amplitude and $\eta$ is the width of the wavepacket in the $x$ direction (perpendicular to the grain boundary). The resulting wavepacket is localized in the $x$ direction, centered around $\mathbf{R}_0$ in real space and around a wavevector $\mathbf{k}_0=k_{x0}\hat{x}+k_{y}\hat{y}$ in reciprocal space. All wavevectors $\mathbf{k}$ are required to be reciprocal lattice vectors of the simulation supercell. With periodic boundary conditions applied in the $y$ direction (parallel to the grain boundary) this gives $k_{y} = 2\pi m/L_y^{\mathrm{sc}}$, where $m$ is an integer and $L_y^{\mathrm{sc}}$ is the size of the supercell in the $y$ direction.

The polarization vectors $\vec{\epsilon}_{j\mathbf{k}}$ and dispersion relation $\omega(\mathbf{k})$ have been obtained from the dynamical matrix of the perfect lattice using the General Utility Lattice Program, GULP\cite{Gale1997GULP1, Gale2003GULP2}, and the constants $A$ and $\eta$ have been set to $0.013$ and $5$ nm, respectively. Since the focus of this study is long-wavelength phonons, the upper limit for $\vert \mathbf{k}_0\vert=\sqrt{k_{x0}^2+k_y^2}$ has been set to $7$ nm$^{-1}$, which limits the possible values of $m$ and $k_{x0}$. To extend the range of possible $m$ values the size of the simulation supercells in the $y$ direction has been increased. For the $9.4^\circ$ boundary it has been tripled, so that $L_y^{\mathrm{sc}} = 4.5$ nm, while for the $17.9^\circ$ boundary it has been doubled, giving $L_y^{\mathrm{sc}} = 4.8$ nm. The supercells of the $9.4^\circ$ and $17.9^\circ$ boundaries are 260 and 400 nm long in the $x$ direction, respectively. Fixed boundary conditions are applied in this direction and all atoms less than $10$ nm from the supercell edge are held immobile. 

In our previous study of phonon scattering at graphene grain boundaries a simple continuum mechanical model of the system was constructed in order to further confirm the results and to facilitate future studies of systems too large to model using MD\cite{Helgee2014flexural}. The model built on the observation that the main cause for scattering of long-wavelength phonons at the grain boundary is the buckling. Here, we have extended the previously used model from one to two dimensions for the case of the $9.4^\circ$ boundary, and incorporated the periodic height variation of the buckling.

 The equations of motion for the displacements are:
\begin{align}\label{eq:contu1}
&\rho \ddot{u}-\partial_x\sigma_{xx}-\partial_y \sigma_{xy} = 0 \\
&\rho \ddot{v}-\partial_x\sigma_{xy}-\partial_y \sigma_{yy} = 0 \\\label{eq:contw1}
&\rho \ddot{w}+\kappa\Delta^2w-\partial_x[\sigma_{xx}\partial_x w+\sigma_{xy}\partial_y w] \\ \nonumber
&-\partial_y[\sigma_{xy}\partial_xw+\sigma_{yy}\partial_yw]=0,
\end{align}
where $u$ is the displacement in $x$, $v$ is the displacement in $y$, $w$ is the out of plane displacement, $\rho$ is the density, $\kappa$ is the bending rigidity and $\sigma_{xx},~\sigma_{xy}$ and $\sigma_{yy}$ are the components of the stress tensor. As in the previous study the grain boundary buckling has been included in the form of a static out-of-plane displacement.

 Finite-difference time-domain methods have been used to propagate wavepackets similar to the ones used in MD and to study scattering against the buckling. Results of these calculations can be directly compared to the MD simulation results. The details of the continuum mechanical model can be found in the Appendix.

\section{Results}
\begin{figure}
  \begin{center}
    \includegraphics[width=\linewidth]{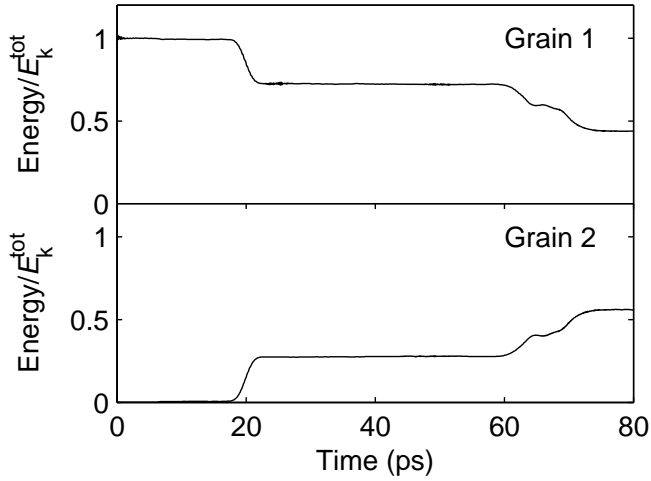}
  \end{center}
  \caption{(Color online) The fraction of the total kinetic energy in grain 1 (top) and grain 2 (bottom) as a function of time for a wavepacket with $k_{x0}=4$ nm$^{-1}$ and $m=2$ scattering at the $9.4^\circ$ boundary. \label{fig:Ekin_t}}
\end{figure}

The time evolution of the kinetic energy in both grains for a wavepacket with $k_{x0}=4$ nm and $m=2$ interacting with the $9.4^\circ$ boundary can be seen in Figure \ref{fig:Ekin_t}. Here, grain 1 is defined as the grain in which the pulse is introduced, and grain 2 is the other grain. Changes in the kinetic energy of the grains can be seen at two points. After $20$ ps, the kinetic energy in grain 1 decreases to $73$~\% of the total kinetic energy while the kinetic energy of grain 2 increases to 27~\%, indicating that the pulse has reached the grain boundary. The second change occurs at $60$ ps, where the energy of grain 1 decreases further in two steps, first to $60$~\% and then to $44$~\%. Between these two points the pulse has been reflected against the fixed boundary conditions, so that the steps at $60$ ps mark the return of the scattered pulses to the grain boundary.

\begin{figure}
  \begin{center}
    \includegraphics[width=\linewidth]{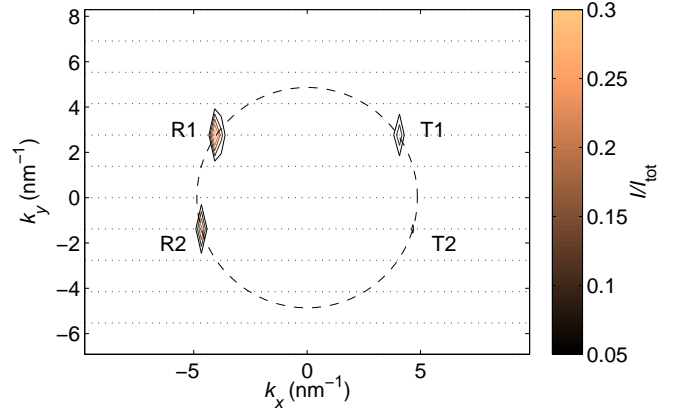}
  \end{center}
  \caption{(Color online) Normalized intensity $I/I_{\mathrm{tot}}$ after scattering at the $9.4^\circ$ boundary ($t=40$ ps) as function of $k_x$ and $k_y$ for a wavepacket with $k_{x0} = 4$ nm$^{-1}$ and $m=2$. R1 and R2 denote the reflected pulses, while T1 and T2 are the transmitted pulses. The dotted lines represent the values of $k_y$ allowed by the boundary conditions and the dashed circle indicates the points with $k_0=\sqrt{k_x^2+k_y^2}$ equal to that of the incident pulse. \label{fig:g1MD}}
\end{figure}

The most surprising feature of Figure \ref{fig:Ekin_t} is the stepwise change in energy beginning at $60$ ps, which seems to indicate that there are two pulses arriving at the grain boundary about $5$ ps apart. A closer examination of the scattered pulses shows that this is indeed the case. Figure \ref{fig:g1MD} shows the intensity of the scattered pulses, normalized by the total intensity, as a function of wavevectors $k_x$ and $k_y$ for $t=40$ ps. Four peaks are seen, two with negative $k_x$, corresponding to reflected pulses, and two transmitted pulses with positive $k_x$. The reflected pulses are labeled R1 and R2. R1 has $k_x=-4$ nm$^{-1}$ and $k_y=2.8$ nm$^{-1}$, while R2 occurs at $k_x= -4.7$ nm$^{-1}$ and $k_y=-1.4$ nm$^{-1}$. Similarly, the transmitted pulses T1 and T2 have $k_x=4,~k_y = 2.8$ nm$^{-1}$ and  $k_x = 4.7,~k_y=-1.4$ nm$^{-1}$, respectively. T1 has the same wavevector as the incident pulse. Since the propagation velocity of the pulse depends on the value of $k_x$, these two pulses will propagate with different velocities and thus give rise to the stepwise change in kinetic energy seen in Figure \ref{fig:Ekin_t}.

\begin{figure}
  \begin{center}
    \includegraphics[width=\linewidth]{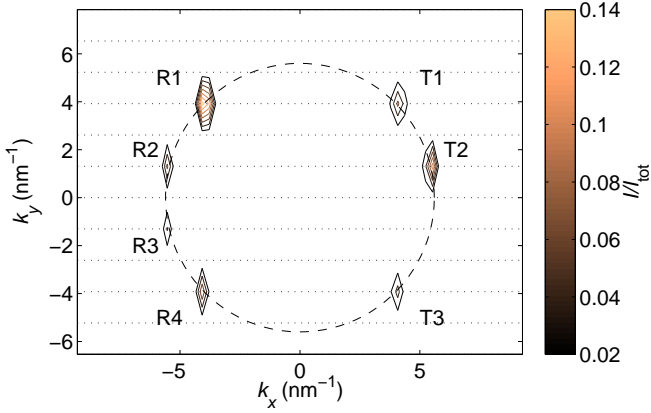}
  \end{center}
  \caption{(Color online) Normalized intensity $I/I_{\mathrm{tot}}$ after scattering at the $17.9^\circ$ grain boundary as function of $k_x$ and $k_y$ for a wavepacket with $k_{x0} = 4$ nm$^{-1}$ and $m=3$. R1 to R4 denote the reflected pulses, while T1, T2 and T3 are the transmitted pulses. The dotted lines represent the values of $k_y$ allowed by the boundary conditions and the dashed circle indicates the points with $k_0=\sqrt{k_x^2+k_y^2}$ equal to that of the incident pulse. \label{fig:gb17int}}
\end{figure}

The same phenomenon is observed at the $17.9^\circ$ grain boundary. Figure \ref{fig:gb17int} shows the normalized intensity after scattering for a pulse with $k_{x0} = 4$ nm$^{-1}$ and $m=3$. Four reflected peaks and three transmitted peaks can be seen. For the reflected peaks, R1 occurs at $k_x = -4.0,~k_y=3.9$, R2 at $k_x = -5.5,~k_y=1.3$, R3 at $k_x = -5.5,~k_y=-1.3$ and R4 at $k_x = -4.0,~k_y=-3.9$ nm$^{-1}$, while the transmitted peaks occur at $k_x = 4.0,~k_y=3.9$ (T1), $k_x = 5.5,~k_y=1.3$ (T2), and $k_x = 4.0,~k_y=-3.9$ nm$^{-1}$ (T3).

Examination of the scattered pulses at both grain boundaries reveal that the difference between the $k_y$ value for the incident pulse, $k_y^{\mathrm{in}}$, and the $k_y$ value for the scattered pulses, $k_y^{\mathrm{sc}}$, can be expressed as
\begin{equation}
k_y^{\mathrm{sc}}-k_y^{\mathrm{in}} = \frac{n2\pi}{L_y},
\end{equation}
where $n$ is an integer and $L_y$ is the grain boundary period. The $k_x$ value of the scattered pulses, $k_x^{\mathrm{sc}}$, is given by momentum conservation:
\begin{equation}
k_x^{\mathrm{sc}}=\sqrt{(k_0^{\mathrm{in}})^2-(k_y^{\mathrm{sc}})^2}.
\end{equation}
This shows that the buckled, periodic grain boundaries act as diffraction gratings for long-wavelength flexural acoustic phonons. Grain boundaries functioning as diffraction gratings for phonons has previoulsy been used to model the behaviour of the thermal conductivity in ionic materials\cite{Omini2000diff}.

 Unlike the previously described case where $m=0$\cite{Helgee2014flexural}, scattering into in-plane vibrational modes is negligible for all cases with $m=1$ and $m=2$. Some movement in the $y$ direction is seen at the $9.4^\circ$ boundary for $m=3$.

\begin{figure}
  \begin{center}
      \includegraphics[width=\linewidth]{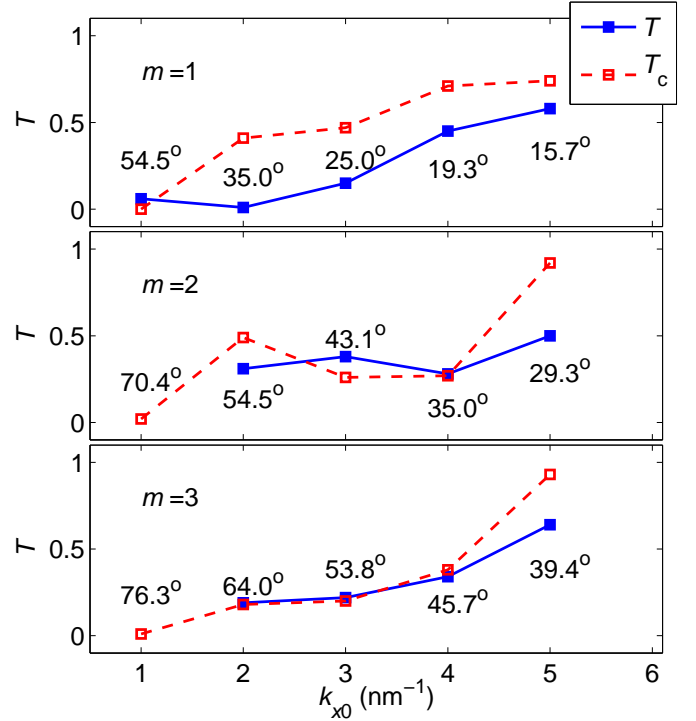}\\
  \end{center}
  \caption{(Color online) Transmission at the $9.4^\circ$ boundary as a function of $k_{x0}$ for $m=1$ (top), $m=2$ (middle) and $m=3$ (bottom). The angle of incidence is indicated beside each data point. Open symbols and dashed lines represent results from the continuum mechanical model. \label{fig:Tvskx}}
\end{figure}

The transmission coefficient $T$ is defined as
\begin{equation}
T=\frac{\langle E_{\mathrm{k}}^{\mathrm{grain 2}}\rangle}{E_{\mathrm{k}}^{\mathrm{tot}}},
\end{equation}
where $E_{\mathrm{k}}^{\mathrm{grain 2}}$ is the kinetic energy in grain 2, $E_{\mathrm{k}}^{\mathrm{tot}}$ is the total kinetic energy and the brackets represent a time average over times between the first scattering at the grain boundary and the time when the first wavepackets reaches the edge of the supercell. Values of $T$ for the $9.4^\circ$ grain boundary for several values of $k_{x0}$ at $m=1,~2$ and $3$ can be seen in Figure \ref{fig:Tvskx}. For all values of $m$, the transmission increases with increasing $k_{x0}$. The increase is monotonic for $m=3$, while for $m=2$ there is a small dip around $k_{x0} = 4$ nm$^{-1}$ and for $m=1$ there is a pronounced trough around $k_{x0} = 2$ nm$^{-1}$. Remarkably, the transmission for $m=1$ and $k_{x0}=2$ nm$^{-1}$  nearly reaches zero, so that no part of the incident pulse is transmitted through the boundary. It can be noted that the dip in the curve for $m=2$ and the trough for $m=1$ occur at the same angle, but for different values of $k_0$. 

\begin{figure}
  \begin{center}
    \includegraphics[width=\linewidth]{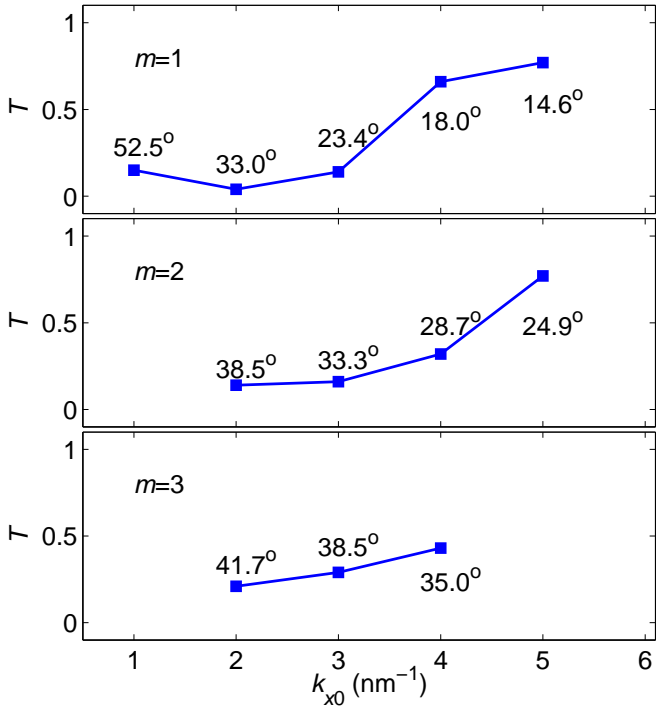}
  \end{center}
  \caption{(Color online) Transmission at the $17.9^\circ$ boundary as a function of $k_{x0}$ for $m=1$ (top), $m=2$ (middle) and $m=3$ (bottom). The angle of incidence is indicated beside each data point. \label{fig:Tvskx17}}
\end{figure}

Figure \ref{fig:Tvskx17} shows the dependence of $T$ on $k_{x0}$ with $m = 1$, $2$ and $3$ for the $17.9^\circ$ boundary. As for the $9.4^\circ$ boundary, the transmission increases with increasing $k_{x0}$. Extremely low transmission is also observed at $m=1$ and $k_{x0}=2$ nm$^{-1}$, corresponding to an incidence angle of $33^\circ$. It is not clear whether there is a minimum at the same angle of incidence for $m=2$, as in the $9.4^\circ$ case, as the transmission is quite low also at slightly larger incidence angles. 

Figure \ref{fig:Tvskx} also contains transmission coefficients $T_{\mathrm{c}}$ obtained from the continuum mechanical model. The qualitative agreement between the continuum mechanical model and the MD results is very good, as the continuum mechanical model clearly reproduces the general trend in the MD data of increasing transmission with increasing $k_{x0}$. The two models agree particularly well for $m=3$, although the continuum mechanical model overestimates the transmission at $k_{x0}=5$ nm$^{-1}$. For $m=2$, the dip around $k_{x0} = 4$ nm$^{-1}$ is reproduced but is wider than in the MD data, extending to $k_{x0}=3$ nm$^{-1}$. The continuum mechanical model also overestimates the transmission at $k_{x0}=5$ nm$^{-1}$. Finally, for $m=1$ the transmission obtained with the continuum mechanical model is higher than that obtained with MD over almost the entire interval. It also does not reproduce the trough at $k_{x0}=2$ nm$^{-1}$, but does reach near-zero values for $k_{x0}=1$ nm$^{-1}$. 

\begin{figure}
  \begin{center}
    \includegraphics[width=\linewidth]{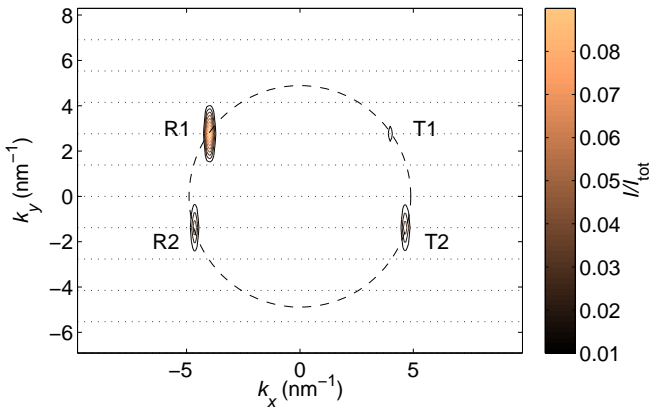}
  \end{center}
  \caption{(Color online) Normalized intensity after scattering as function of $k_x$ and $k_y$ for a wavepacket with $k_{x0} = 4$ nm$^{-1}$ and $m=2$, from the continuum mechanical model. The dotted lines represent the values of $k_y$ allowed by the boundary conditions and the dashed circle indicates the points with $k_0=\sqrt{k_x^2+k_y^2}$ equal to that of the incident pulse.\label{fig:g1CM}}
\end{figure}

In addition to the transmission coefficient, the continuum mechanical model should reproduce the diffraction seen in MD. Figure \ref{fig:g1CM} shows the intensity obtained from the continuum mechanical model after scattering as function of $k_x$ and $k_y$ for $k_{x0} = 4$ and $m=2$, corresponding to the MD results presented in Figure \ref{fig:g1MD}. It is clear that the same peaks appear, showing that diffraction occurs also in the continuum mechanical model. Compared to the MD results T1 appears to be underestimated and T2 overestimated, possibly due to that the model of the boundary buckling used in the continuum mechanical model does not reproduce the actual curvature of the grain boundary buckling in sufficient detail.

\section{Conclusion}
In summary, the effects of the angle of incidence on the scattering of long-wavelength flexural phonons against grain boundaries in graphene have been studied using molecular dynamics. The considered grain boundaries, two buckled symmetric tilt grain boundaries with misorientation angles $9.4^\circ$ and $17.9^\circ$, have been found to act as diffraction gratings for long-wavelength flexural phonons. In addition, near-zero transmission has been observed for angles near $35^\circ$ and small wavevector magnitudes. A continuum mechanical model of the system containing the $9.4^\circ$ boundary has been constructed and shown to qualitatively agree with the MD results, giving insights into the scattering mechanism and providing a starting point for studies of systems too large to be modeled atomistically. The presented results improve our understanding of how phonons interact with grain boundaries in graphene and suggest that such defects could indeed be useful in manipulating the vibrational properties of the material.

\appendix*
\section{Continuum mechanical modeling}
\label{app:A}
In the continuum mechanical model the graphene sheet is described as a thin plate. The equations of motion for the displacements are 
\begin{align}\label{eq:contu}
&\rho \ddot{u}-\partial_x\sigma_{xx}-\partial_y \sigma_{xy} = 0,\\ \label{eq:contv}
&\rho \ddot{v}-\partial_x\sigma_{xy}-\partial_y \sigma_{yy} = 0, \\\label{eq:contw}
&\rho \ddot{w}+\kappa\Delta^2w-\partial_x[\sigma_{xx}\partial_x w+\sigma_{xy}\partial_y w] \\ \nonumber
&-\partial_y[\sigma_{xy}\partial_xw+\sigma_{yy}\partial_yw]=0,
\end{align}
where $u$ is the displacement in $x$ (perpendicular to the boundary), $v$ is the displacement in $y$ (parallel to the boundary), $w$ is the out-of-plane displacement, $\kappa$ is the bending rigidity, $\rho$ is the density and $\sigma_{xx},~\sigma_{xy}$ and $\sigma_{yy}$ are the elements of the two-dimensional stress tensor. To model the grain boundary buckling a static out-of-plane displacement $w_0(x,y)$ is introduced. The introduction of this out-of-plane displacement gives rise to static displacements in the in-plane directions, so that the total displacements must be written
\begin{align}
u(x,y,t) = u_0(x,y)+u_1(x,y,t) \\
v(x,y,t) = v_0(x,y)+v_1(x,y,t) \\
w(x,y,t) = w_0(x,y)+w_1(x,y,t) 
\end{align}
where $u_1(x,y,t),~v_1(x,y,t)$ and $w_1(x,y,t)$ are the time-dependent displacements. However, the displacements determine the stress tensor components through the relations 
\begin{align}
  \sigma_{xx}=&(\lambda+2\mu)\left[\partial_xu+\frac{(\partial_xw)^2}{2}\right]+\lambda\left[\partial_y v+\frac{(\partial_y w)^2}{2}\right] \\ \nonumber
  \sigma_{yy}=&\lambda\left[\partial_x u+\frac{(\partial_x w)^2}{2}\right]+(\lambda+2\mu)\left[\partial_yv+\frac{(\partial_yw)^2}{2}\right] \\ \nonumber
  \sigma_{xy} =& \mu\left[\partial_xv+\partial_yu+\partial_xw\partial_yw\right],
\end{align}
where $\lambda$ and $\mu$ are Lam\'{e} parameters. Thus, the stress tensor elements can also be divided into a time-dependent term $\sigma^1_{ij}$ ($i,j=x,y$) and a time-independent term $\sigma^0_{ij}$. The two-dimensional stress tensor components $\sigma^0_{xx}$, $\sigma^0_{yy}$ and $\sigma^0_{xy}$ are related through the Airy stress function\cite{LL} $\chi$, such that
\begin{equation} \label{eq:airy}
  \sigma_{xx}^0 = \frac{\partial^2 \chi}{\partial y^2}, \quad \sigma_{yy}^0 = \frac{\partial^2 \chi}{\partial x^2}, \quad \sigma_{xy}^0 = -\frac{\partial^2 \chi}{\partial x\partial y}.
\end{equation}
It follows that the time-independent terms of the stress tensor will vanish in Equation \ref{eq:contu} and \ref{eq:contv}, but not in Equation \ref{eq:contw}. The equations of motion for the time-dependent displacements thus become
\begin{align}
  &\rho\ddot{u}_1-\partial_x\sigma_{xx}^1-\partial_y\sigma_{xy}^1=0 \\
  &\rho\ddot{v}_1 -\partial_x\sigma_{xy}^1-\partial_y\sigma_{yy}^1 = 0 \\
  &\rho\ddot{w}+\kappa\Delta^2(w_0+w_1)- \\ \nonumber
 &\partial_x\left[(\sigma_{xx}^0+\sigma_{xx}^1)\partial_x(w_0+w_1)\right. \\ \nonumber
 & \left.+(\sigma_{xy}^0+\sigma_{xy}^1)\partial_y(w_0+w_1)\right]- \\ \nonumber
 &\partial_y\left[(\sigma_{xy}^0+\sigma_{xy}^1)\partial_x(w_0+w_1) \right. \\ \nonumber
 & \left.+(\sigma_{yy}^0+\sigma_{yy}^1)\partial_y(w_0+w_1)\right] = 0.
 \end{align}
When solving these equations, any terms that are not linear in the derivatives of $u_1(x,t)$, $v_1(x,t)$ or $w_1(x,t)$ can be ignored due to small vibrational amplitudes. 

Finite-difference time-domain methods have been used to solve Equations \ref{eq:contu}-\ref{eq:contw}. As in our previous paper\cite{Helgee2014flexural}, the equations have been discretized using standard discretization schemes\cite{numrec} with step sizes $\Delta x=\Delta y=0.05$ nm and $\Delta t = 0.4\sqrt{dx^4/4\kappa}=0.8$ fs. The Lam\'{e} parameters, bending rigidity and density have been set to the values given by the modified Tersoff potential, i.e., $\mu = 167$ N m$^{-1}$, $\lambda = 23$ N m$^{-1}$, $\kappa = 2.8\times10^{-19}$ J and $\rho = 7.42\times 10^{-7}$ kg m$^{-2}$. Fixed boundary conditions are applied in the $x$ direction and periodic boundary conditions are applied in the $y$ direction. The initial conditions are  
\begin{align}
w_{1}(x,y,t=0) = \mathrm{Re}\left[\sum_{k}a_{\mathbf{k}}e^{i(\mathbf{k}\cdot \mathbf{R}-\omega(k_0)t)}\right], \\ \nonumber
\partial_t w_{1}(x,y,t=0)=\mathrm{Re}\left[-i\omega(k_0)\sum_{k}a_{\mathbf{k}}e^{i(\mathbf{k}\cdot \mathbf{R}-i\omega(k_0)t)}\right],
\end{align}
with
\begin{equation}
a_{\mathbf{k}} = Ae^{-\eta^2(k_x-k_{0x})^2}e^{-i\mathbf{k}\cdot\mathbf{R}_0}.
\end{equation}
As in the MD simulations, $\mathbf{k} = k_{x}\hat{x}+k_y\hat{y}$ is a wavevector allowed by the boundary conditions, $\mathbf{R}=x\hat{x}+y\hat{y}$ is a position, $A=0.01$ nm is an amplitude and $\eta = 2$ nm is the width of the wavepacket. The wavepacket is centered around $\mathbf{R}_0$ in real space and $\mathbf{k_0} = k_{x0}\hat{x}+k_y\hat{y}$ in reciprocal space, and $\omega(k_0)$ is the frequency of out-of-plane vibrations with wavevector $k_0$.  

The static out-of-plane displacement is set to
\begin{equation}
w_0(x,y) = A_{\mathrm{b}} e^{-x^2/2\xi^2}\left(1+a\sin\left(\frac{2\pi m y}{L_y}\right)\right)
\end{equation}
where $L_y$ is the system size in the $y$ direction. Fitting to the shape of the buckling of the $9.4^\circ$ boundary produced by MD simulations gives $A_{\mathrm{b}} = 0.55$ nm, $\xi = 0.72$ nm and $a = 0.01$. As in the MD simulations $L_y = 4.5$ nm, so $m$ must be set to $3$ to obtain the correct periodicity in $y$. The system length in the $x$ direction, $L_x$, is set to $100$ nm. 

\begin{figure}
\begin{center}
\subfigure[]
{\includegraphics[width=\linewidth]{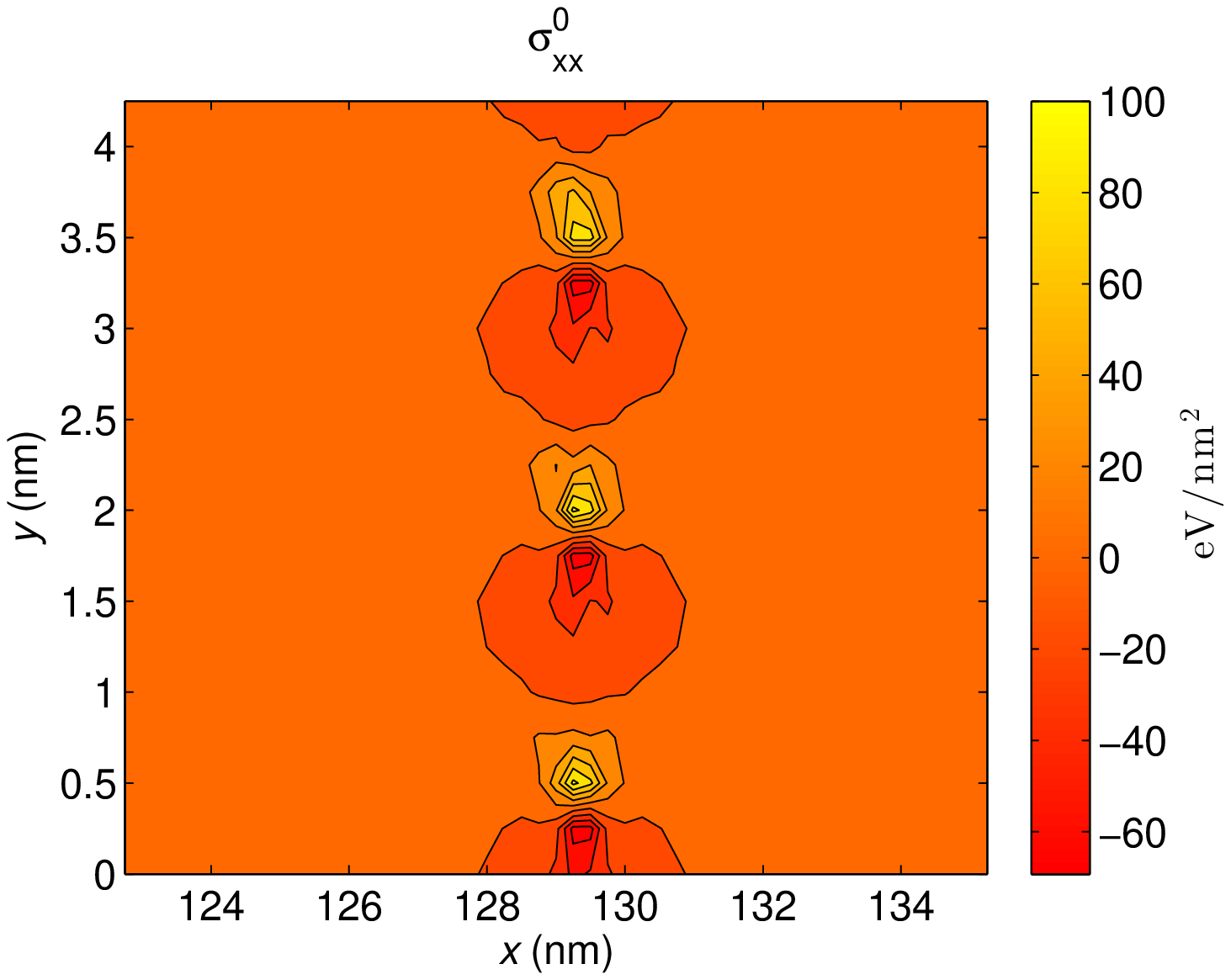}\label{subfig:sigmaxxMD}}
\qquad
\subfigure[]
{\includegraphics[width=\linewidth]{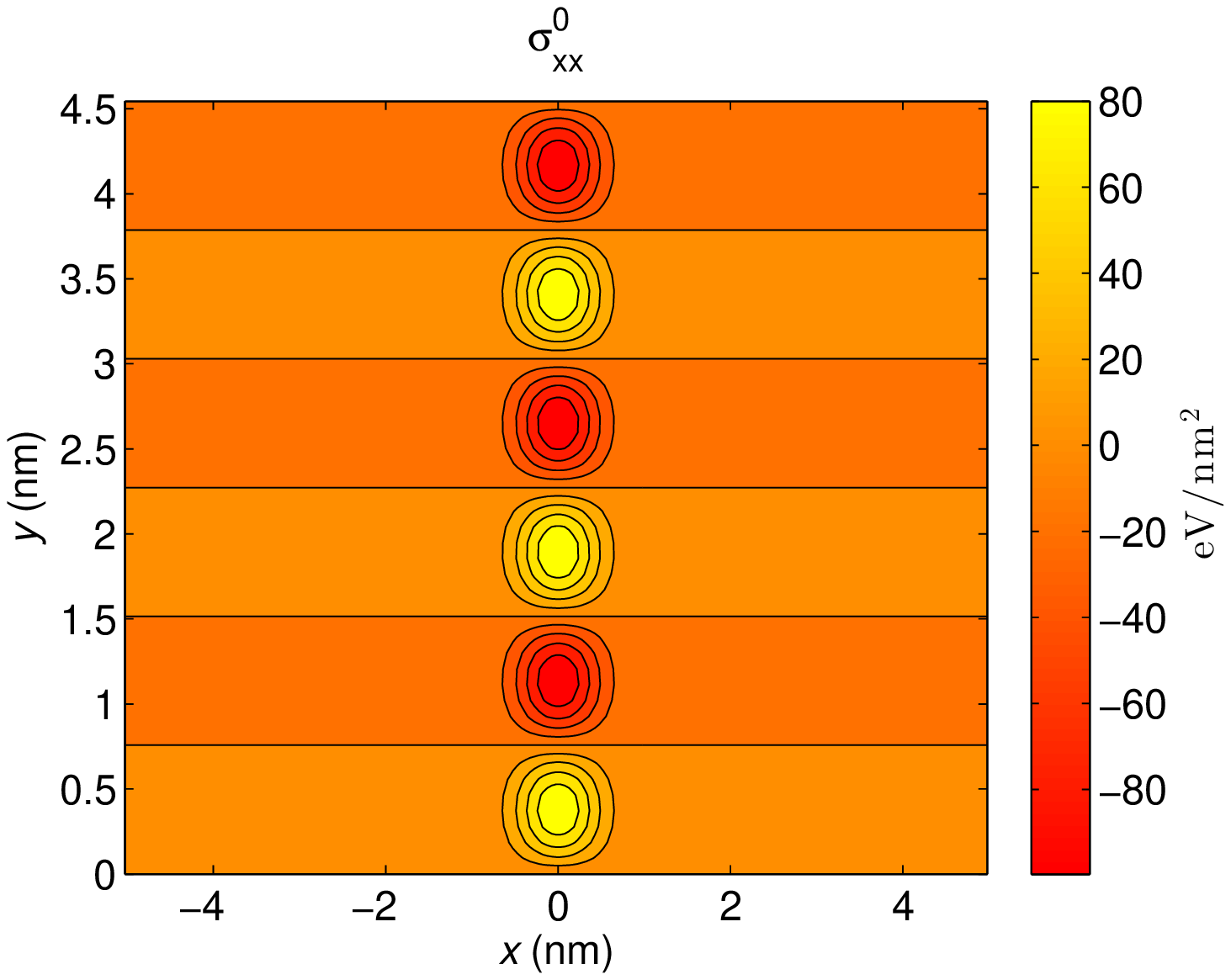}\label{subfig:sxxCM}}
\caption{(Color online) The stress tensor component $\sigma^0_{xx}$ close to the grain boundary (a) as obtained from MD and (b) as approximated according to Equation \ref{eq:sxx}. Note that the grain boundary is located at $x=129$ nm in the MD simulatons and at $x=0$ in the continuum mechanical model.}
\label{fig:sxx}
\end{center}
\end{figure}

In addition to the static out-of-plane displacement, the time-independent terms in the stress tensor components are also needed. These have been obtained by fitting to the (approximate) stress tensor components obtained from MD. Starting with $\sigma^0_{xx}$, it is seen that if we set
\begin{equation}\label{eq:sxx}
\sigma^0_{xx} = e^{-2x^2/\xi^2}\sin\left(\frac{2 \pi m y}{L_y}\right)
\end{equation}
we obtain a good qualitative correspondence to the MD data (see Figure \ref{fig:sxx}). 

\begin{figure}
\begin{center}
\subfigure[]
{\includegraphics[width=\linewidth]{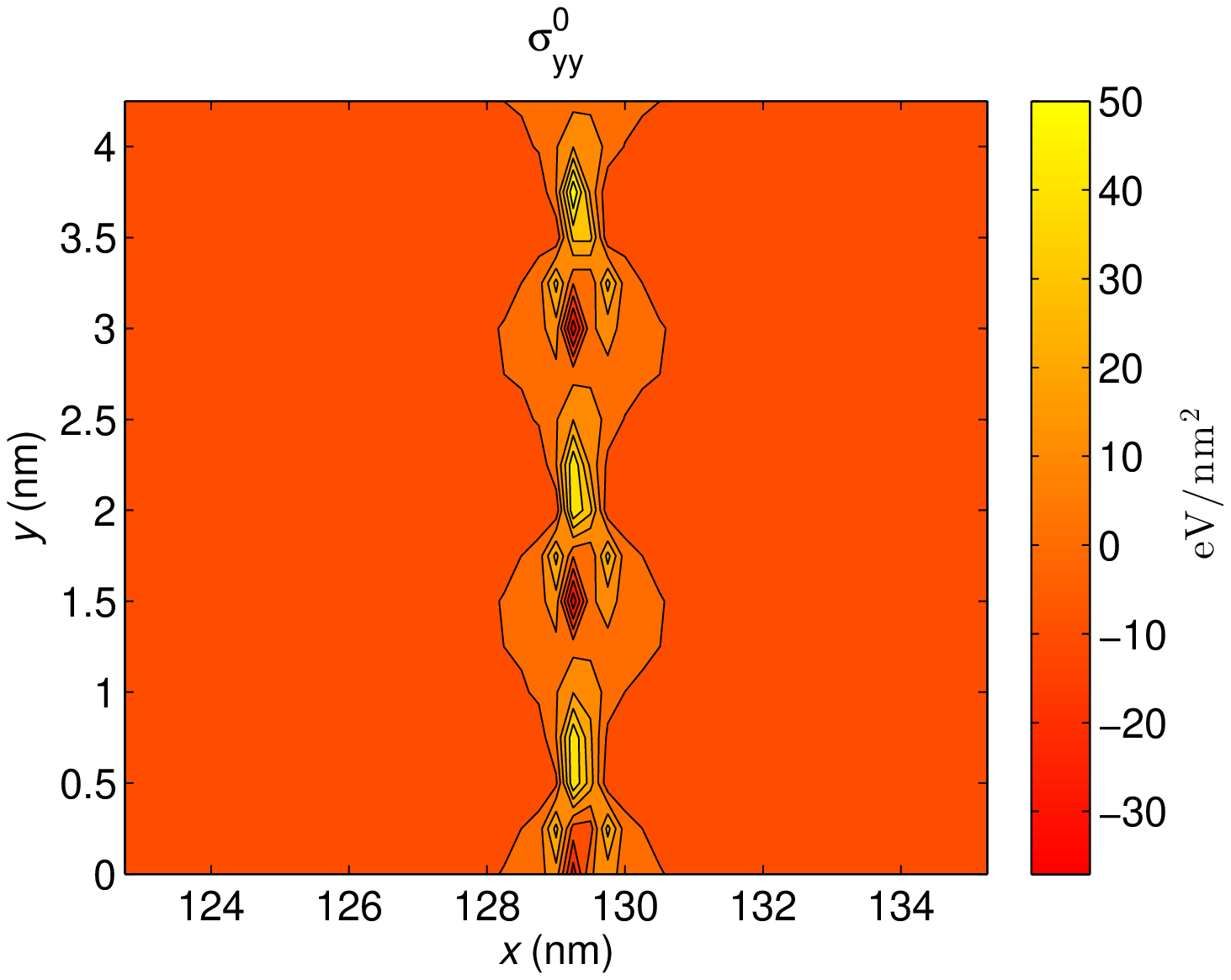}\label{subfig:sigmayyMD}}
\qquad
\subfigure[]
{\includegraphics[width=\linewidth]{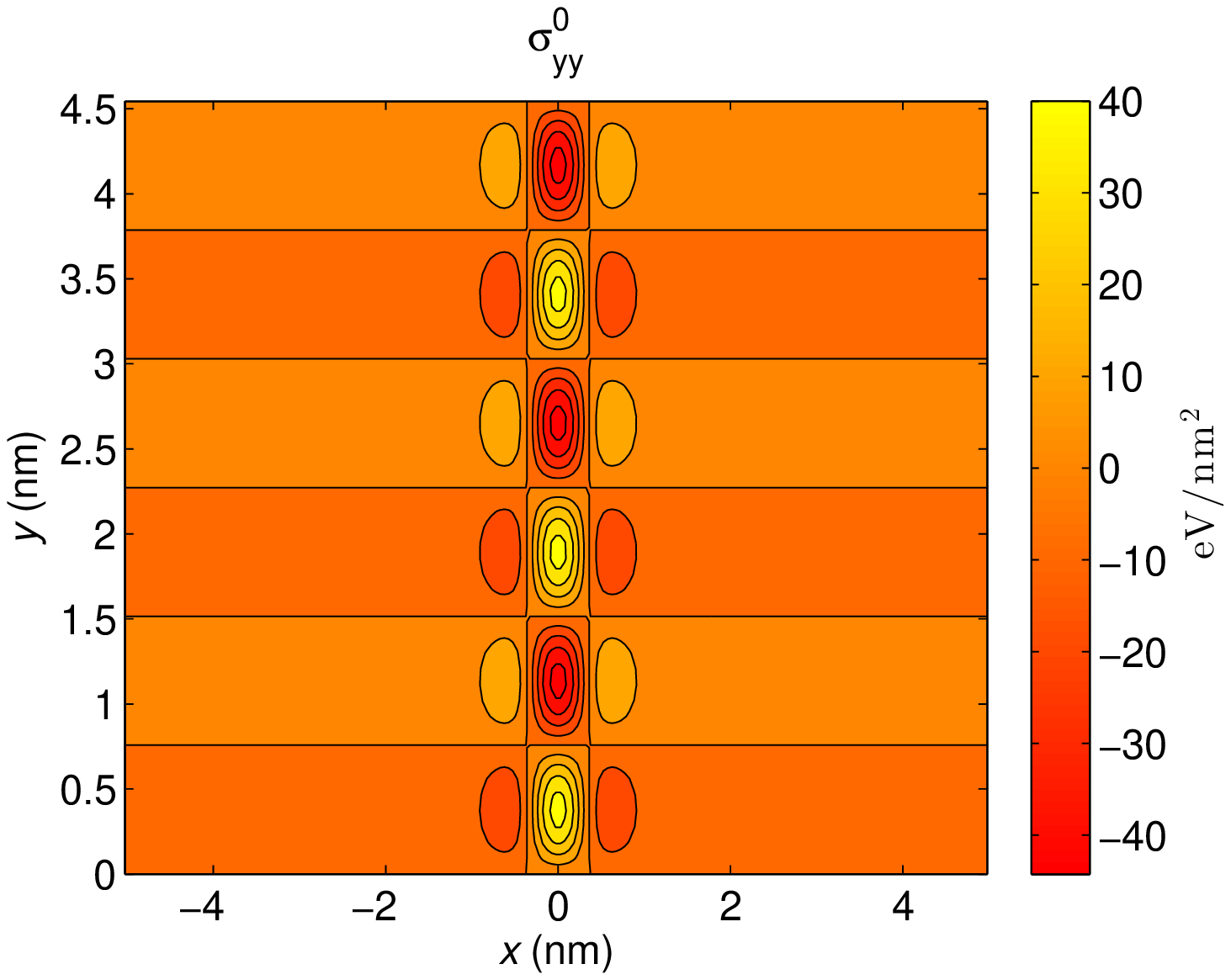}\label{subfig:syyCM}}
\caption{(Color online) The stress tensor component $\sigma^0_{yy}$ (a) as obtained from MD and (b) as approximated according to Equation \ref{eq:syy}.}
\label{fig:syy}
\end{center}
\end{figure}

\begin{figure}
\begin{center}
\subfigure[]
{\includegraphics[width=\linewidth]{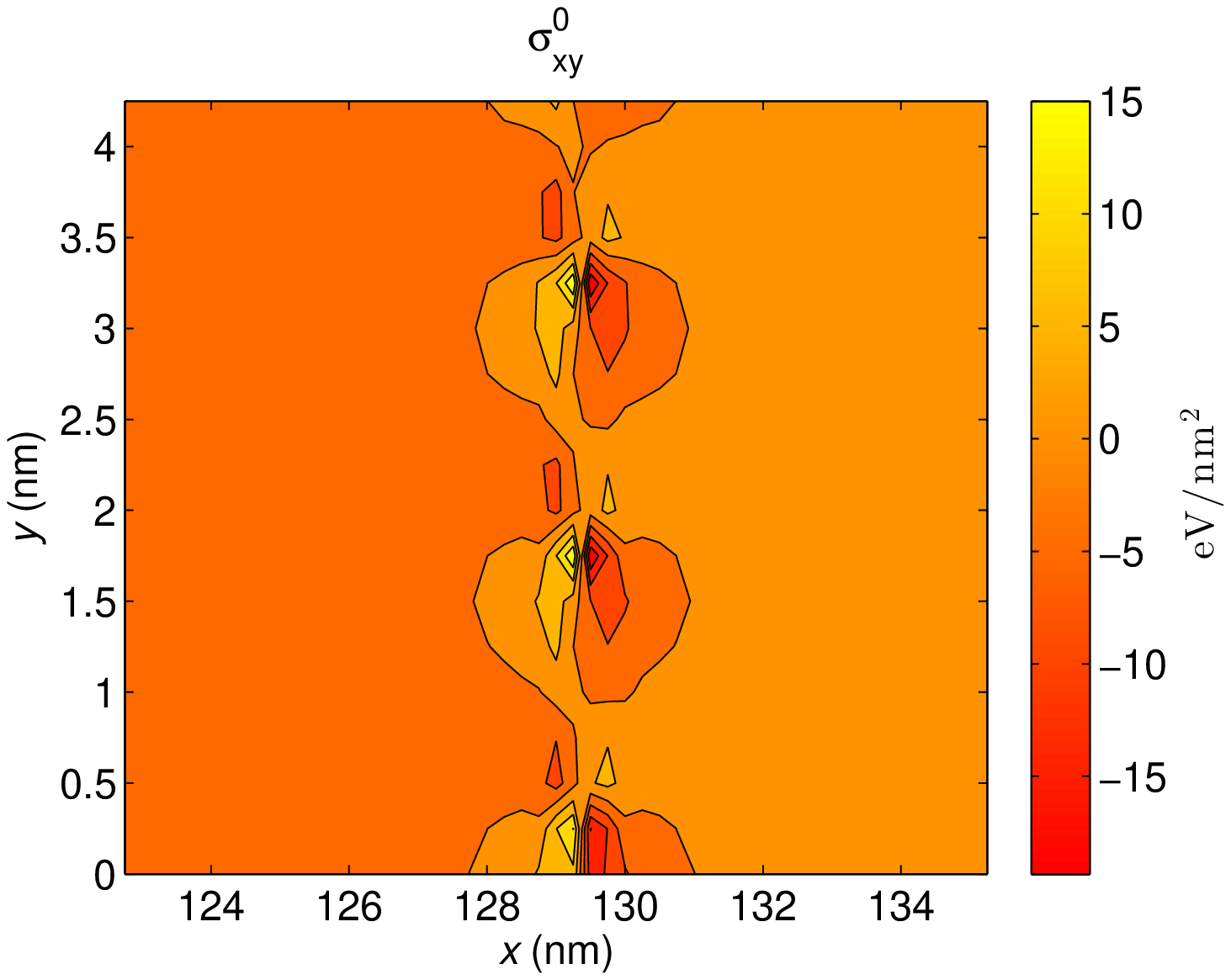}\label{subfig:sigmaxyMD}}
\qquad
\subfigure[]
{\includegraphics[width=\linewidth]{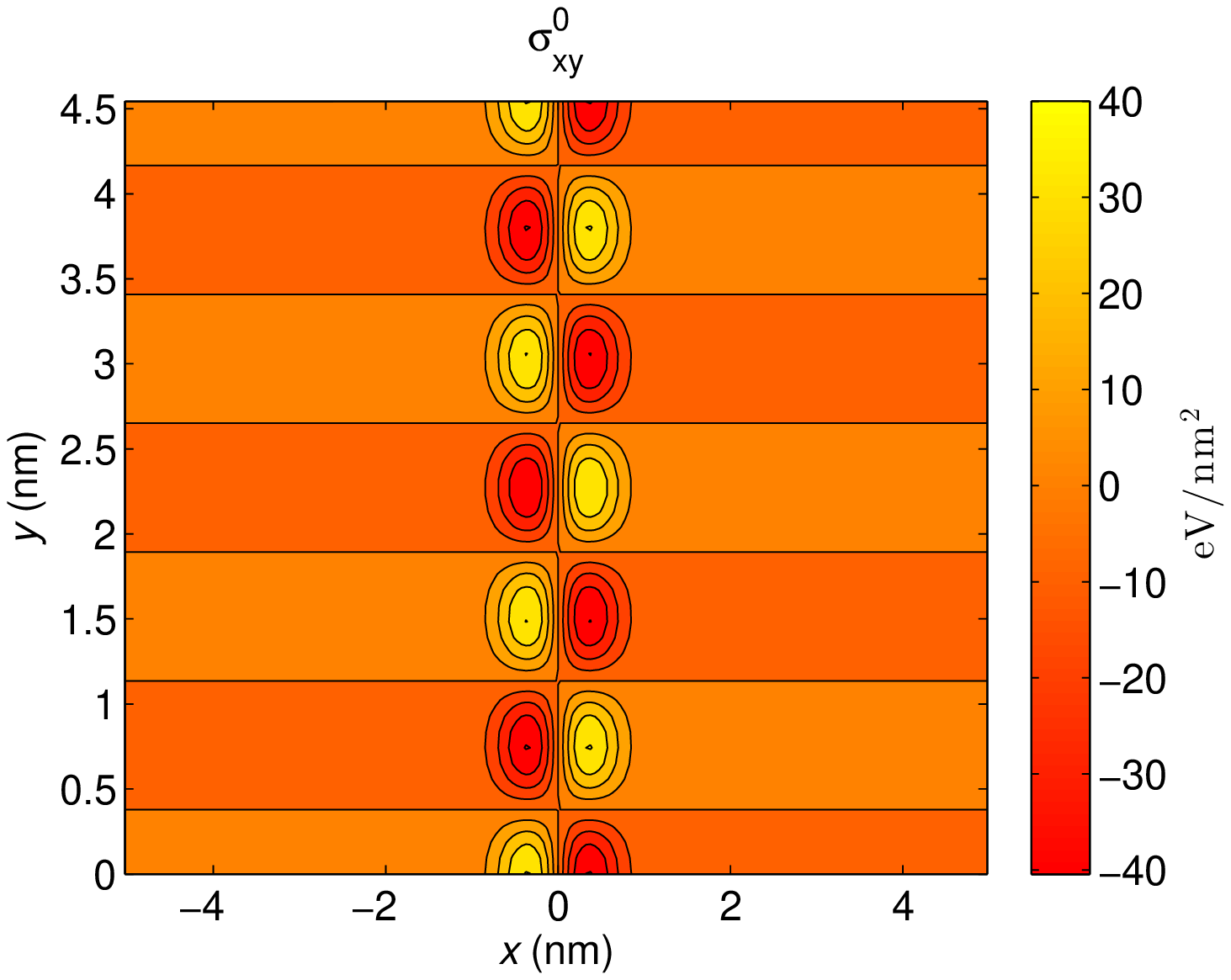}\label{subfig:sxyCM}}
\caption{(Color online) The stress tensor component $\sigma^0_{xy}$ (a) as obtained from MD and (b) as approximated according to Equation \ref{eq:sxy}.}
\label{fig:sxy}
\end{center}
\end{figure}

To satisfy the relations between the stress tensor components given by Equation \ref{eq:airy}, we must then set
\begin{align}\label{eq:syy}
&\sigma^0_{yy} = -\left(\frac{L_y}{2\pi m}\right)^2\left(-\frac{4}{\xi^2}+\frac{16x^2}{\xi^4}\right)e^{-2x^2/\xi^2}\sin\left(\frac{2 \pi m y}{L_y}\right)\\ \label{eq:sxy}
&\sigma^0_{xy} = -\left(\frac{L_y}{2\pi m}\right)\frac{4x}{\xi^2}e^{-2x^2/\xi^2}\cos\left(\frac{2 \pi m y}{L_y}\right).
\end{align}
As can be seen in Figures \ref{fig:syy} and \ref{fig:sxy}, this functional form of the stress tensor components does reproduce the MD result in a qualitiative manner.

In order to compare the results of the continuum mechanical model to those obtained from MD, the transmission coefficient $T_{\mathrm{c}}$ is calculated as
\begin{widetext}
\begin{equation}
T_{\mathrm{c}} = \left\langle \frac{\Delta x \rho \sum_{x_i>0}(\omega_x^2u_1^2(x_i,y_j,t_n)+ \omega_y^2v_1^2(x_i,y_j,t_n) + \omega_z^2w_1^2(x_i,y_j,t_n))}{2E^{\mathrm{tot}}} \right\rangle
\end{equation}
\end{widetext}
where $\omega_x$, $\omega_y$ and $\omega_z$ are the frequencies of vibrations in the $x$, $y$ and $z$ directions, $x_i=i\Delta x$ and $y_j=j\Delta y$ indicate a point on the discretization grid, and $t_n=n\Delta t$ is the timestep. The time average is taken over times after scattering against the static out-of-plane displacement and the total energy $E^{\mathrm{tot}}$ is given by
\begin{widetext}
\begin{equation}
E^{\mathrm{tot}} = \frac{\Delta x \rho}{2} \sum_{x_i=-L_x/2}^{L_x/2}(\omega_x^2u_1^2(x_i,y_j,t_n) + \omega_y^2v_1^2(x_i,y_j,t_n) + \omega_z^2w_1^2(x_i,y_j,t_n)).
\end{equation}
\end{widetext}

\section*{Acknowledgements}
The authors would like to thank Prof. Jari Kinaret for rewarding discussions. We also acknowledge financial support from the Swedish Research Council (VR) and the EU Graphene Flagship (grant no. 604391).

\end{document}